\documentclass[trackchanges, twocolumn]{aastex7}

\usepackage{amsmath}

\begin{document}

\title{
No Evidence for Second-Scale Periodicity in FRB 20201124A from FAST Observations
}

\correspondingauthor{Dotan Gazith}

\author[0000-0001-6698-3693,sname=Gazith,gname=Dotan]{Dotan Gazith}
\affiliation{Department of Particle Physics and Astrophysics, Weizmann Institute of Science, 76100 Rehovot, Israel}
\email{dotan.gazith@weizmann.ac.il}

\author[0000-0001-5162-9501, sname=Zackay, gname=Barak]{Barak Zackay}
\affiliation{Department of Particle Physics and Astrophysics, Weizmann Institute of Science, 76100 Rehovot, Israel}
\email{barak.zackay@weizmann.ac.il}

\begin{abstract}

Fast Radio Bursts (FRBs) are bright and short radio flashes of cosmological origin. Although a great number of FRBs were detected in the last two decades, their progenitors and the physical processes that create them are unknown. In recent years, magnetars were proposed as one of the leading progenitor candidates. A striking feature that can hint at such magnetar origin is second-scale periodicity.
In this paper we define a robust procedure to search for such periodicity, and estimate the significance of its results. We search for such periodicity in the bursts of FRB 20201124A, observed by the Five-hundred-meter Aperture Spherical Telescope (FAST) between April and June 2021. 
Our analysis does not find any significant periodicity.
We discuss the differences between our non-detection and the $\sim1.7\text{ s}$ periodicity claim by \cite{du_second-scale_2025}.

\end{abstract}

\section{Introduction}\label{sec:intro}

Fast radio bursts (FRBs) are ms-duration radio flashes, typically observed in the GHz frequencies, originating from cosmological distances, corresponding to the redshift range $z\sim0.1-1$.
Since the first discovery of an FRB \cite{lorimer_bright_2007} a diverse phenomenology has emerged, including a great range of repetition rates \citep{xu_fast_2022, feng_extreme_2023}, day-scale periodicity \citep{chimefrb_collaboration_periodic_2020, cruces_repeating_2021, rajwade_possible_2020}, and association with various environments, such as various types of galaxies \citep{tendulkar_host_2017, xu_fast_2022}, persistent radio sources \citep{marcote_repeating_2017, marcote_repeating_2020, niu_repeating_2022}, some showing extremely young age \citep{waxman_origin_2017}, globular clusters \citep{kirsten_repeating_2022} hinting at older populations, and the galactic extremely weak FRB emitted by a magnetar \citep{zhang_highly_2020, bochenek_stare2_2020, scholz_bright_2020, chimefrb_collaboration_bright_2020, kirsten_detection_2021, bochenek_fast_2020}.

Despite this wealth of observational data, a consensus regarding the systems and mechanisms generating FRBs has not been achieved yet.
Motivated by the galactic FRB emitted by a magnetar, and the family of magnetar origin of FRBs, second scale periodicity has been searched for in bursts' arrival times \citep{niu_fast_2022, du_thorough_2024}, with a recent claimed detection by \cite{du_second-scale_2025}.

In this paper we rigorously define the search procedure, and determine its parameters and significance estimation process. We execute the search for periodicity on the bursts of FRB 20201124A, that were detected by the Five-hundred-meter Aperture Spherical Telescope (FAST) between April and June 2021 \citep{xu_fast_2022}.

In contrast with the findings of \cite{du_second-scale_2025}, we do not find a significant periodicity in MJDs 59310 and 59347. 
We discuss the difference between the procedures, and independently verify that, even when using their test statistic, the found periodicity is insignificant. For MJD 59310, the primary difference is our family-wise error rate estimation, while for MJD 59347, the difference is also rejecting two triplets of extremely close bursts.

\section{Data}
Between April and June 2021, \cite{xu_fast_2022} detected a large number of bursts from FRB 20201124A using the 19-beam receiver of the FAST, spanning 1.0 GHz to 1.5 GHz. We use the barycentric arrival times reported by \cite{xu_fast_2022}, containing 1863 bursts between 59307-59360 MJD.

\section{Methods}
\subsection{Periodicity search}
\subsubsection{Time of Arrival Pre-Processing}
\label{subsubsec:preprocess}
\begin{figure}
    \centering
    \plotone{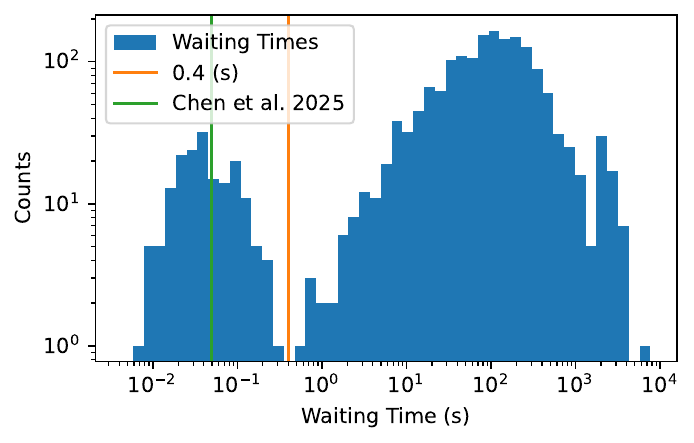}
    \caption{Waiting time distribution of FRB 20201124A. Vertical line indicate the thresholds used to separate between the short and long modes of the distribution in this work (orange) and by \cite{du_second-scale_2025} (green).}
    \label{fig:wait_times}
\end{figure}

The waiting time distribution was studied for different rFRBs, showing a clear deviation from the standard Poisson process naive assumption. The waiting time distribution in this data set, as shown in Figure \ref{fig:wait_times}, is multi-modal, with short ($\sim50\text{ ms}$) and long ($\sim100\text{ s}$) modes. For the purpose of the periodicity search, we ignore the bursts from the short mode, by throwing away bursts with a burst up to $0.4\text{ s}$ before them. As seen in Figure \ref{fig:wait_times}, this threshold separates well between the short and long modes.

This thresholding is motivated by the magnetar progenitor model, in which the short mode is associated with some substructure in the magnetar. In Appendix \ref{app:stat_in_short} we also show that including the short mode significantly changes the search's distribution.

\subsubsection{Timing Model}
We search for a constant frequency timing model for the bursts arriving each day separately (spread over 2-4 hours a day).
\begin{equation}
\label{eq:timing_model}
    \phi_i = ft_i \text{\;mod\;} 1 \, ,
\end{equation}
where $t_i$ and $\phi_i$ are the $i$'th arrival time and phase residual and $f$ is the frequency search for. We search for frequencies in the range $0.1-10\text{ Hz}$.

We do not consider frequency derivatives, because only characteristic ages shorter than roughly $3\text{yr}\left(\frac{f}{1\text{Hz}}\right)\left(\frac{T}{10^4\text{s}}\right)^2$ are detectable in a single observation of time $T$.
We choose to neglect timing models including binary orbits, keeping them out of this paper's scope, but more advanced algorithms, as discussed in \cite{gazith_recovering_2025} can be used to overcome the run-time challenge those additional parameters present.

\subsubsection{Test Statistic}
\label{subsubsec:test_stat}
The criterion for deciding if a given period is significant or not has two parts. The first is computing a scalar called "test statistic" that the higher it is, the more significant the period is. The second is the exact determination of the distribution of the test statistic, identifying the detection threshold.

We propose the use of the H-test \citep{de_jager_powerful_1989} as a test statistic, with some changes due to the low number of TOAs. We will also consider the binned $\chi^2$ test, to maintain compatibility with \cite{du_second-scale_2025}. 

In the $H$ test, harmonic sums of different orders are considered, and the statistic is the maximum among the corrected harmonic sums.
\begin{subequations}
\begin{equation}
    \hat{\alpha}_k = \sum_{i=1}^n\cos k\phi_i \: ;\:
    \hat{\beta}_k = \sum_{i=1}^n\sin k\phi_i
\end{equation}
\begin{equation}
    Z_m^2 = \frac{2}{n}\sum_{k=1}^m \left(\hat{\alpha}_k^2 + \hat{\beta}_k^2\right)
\end{equation}
\begin{equation}
    H_N = \max_{1\leq m\leq N}\left(Z_m^2 - 4m+4\right)\,.
\end{equation}
\end{subequations}
A typical choice is $N=20$, but, motivated by the similarity between an $H_N$ test and a $\chi^2$ test with $2N$ bins (due to Parseval's theorem, after binning) and the previous discussion regarding the bad $\chi^2$ approximation, we choose $N=5$.

In the $\chi^2$ test, as performed by \cite{du_second-scale_2025}, the phase residuals are binned to $n=30$ bins, and the deviations from the expected number of bins are summed in quadrature
\begin{equation}
\label{eq:chi2}
    \chi^2=\sum_{j=1}^n\frac{(O_j-E_j)^2}{(n-1)E_j}\,,
\end{equation}
where $O_j$ and $E_j$ are the observed  and expected number of bursts in the $j$-th bin. It should be noted that because in most days the total number of bursts is only a few tens, $E_j\leq2$, hence the normal approximation doesn't hold and the statistic does not follow the $\chi^2$ distribution.

\subsubsection{Frequency Spacing}
\label{subsubsec:freq_spacing}
To ensure a thorough search over the frequency range, proper frequency sampling is required. The test statistic is calculated using the folded phases $\phi_i$, therefore, the frequency resolution is directly set by its effect on the folded phase. Using the timing model, as defined in Equation \ref{eq:timing_model}, we get
\begin{equation}
    \Delta f \sim \frac{\Delta \phi}{\max t_i - \min t_i}\,,
\end{equation}
where $\Delta\phi$ is the phase resolution, as derived from the test statistic.
For the $H_5$ test we chose, phase resolution is of order $10\%$, and we choose $\Delta\phi=1\%$ to ensure completeness at the expense of unnecessary computation (which isn't a bottleneck in this search). For the $\chi^2$ test, there is no fixed phase resolution, due to the discrete nature of the binning operation (which also requires enumeration over the best global sub-bin offset). To properly reconstruct the results of \cite{du_second-scale_2025}, we use their $1\times10^{-7}\text{ s}$ period spacing \footnote{Private communication.}.

\subsection{Significance determination}
Given the low number of bursts in each day (relative to the number of bins and harmonics), we do not use the theoretical distributions, as deriving them requires invalid approximations. We empirically estimate the significance by sampling sets of TOAs for each day, searching for periodicity in them, and keeping the maximum from each search.
TOAs are sampled in two ways:
\begin{enumerate}
    \item Uniformly in the range of the detected bursts.\footnote{In most days, observations were not continuous \citep{xu_fast_2022}, but we neglect this effect as we don't know exactly the observation windows, and expect this to have negligible effect on the results.}
    \item By sampling waiting times from the long log-normal mode, using $\text{LogNormal}(4.5, 1.4^2)\text{ s}$, ensuring the total duration is similar to that of the detected bursts up to $10\%$ using rejection sampling.\footnote{The total duration criterion can't be satisfied for MJD $59315$, as the burst rate in it is too high. Therefore only uniform sampling is used for this day.}
\end{enumerate}
In both ways we keep the number of bursts in a day the same, and ensure no two bursts are closer than the preprocessing threshold $0.4\text{ s}$.
We sample $500$ sets of TOAs for each day.

\section{Results}
The dataset we analysed contains 1863 bursts, out of which 1690 passed our pre-processing threshold. Out of the 45 days analysed, using 2 test statistics and 2 significance determination sampling strategies, only MJD 59310 passes the bar of its 500 searches on sampled data, when analysed using the $\chi^2$ test and log-normal sampling strategy. To improve the significance estimate of this result, we fit a Gumbel distribution\footnote{Gumbel distribution is the extreme-value distribution of the $\chi^2$ distribution, which is the motivation for this choice.} to the samples and calculated the p-value of the search result assuming the fitted Gumbel distribution
\begin{equation*}
    \text{p-value}_{59310}\equiv P(\chi^2_{\text{search}}\le\chi^2_{\text{Gumbel}})=10^{-3} \, .
\end{equation*}
This p-value needs to be compared to the typical $5\sigma$ threshold, which corresponds to a false-alarm probability
\begin{equation*}
    \alpha \sim 2\times10^{-7}\,,
\end{equation*}
then corrected for the multiple days, tests and sampling strategies considered, for example using the Bonferroni correction, results in the threshold
\begin{equation*}
    \frac{\alpha}{m}\sim10^{-9}\,.
\end{equation*}
Comparing the recovered p-value and threshold for detection results in the clear conclusion that our search result for MJD 59310 is not significant.

\begin{figure}
    \centering
    \plotone{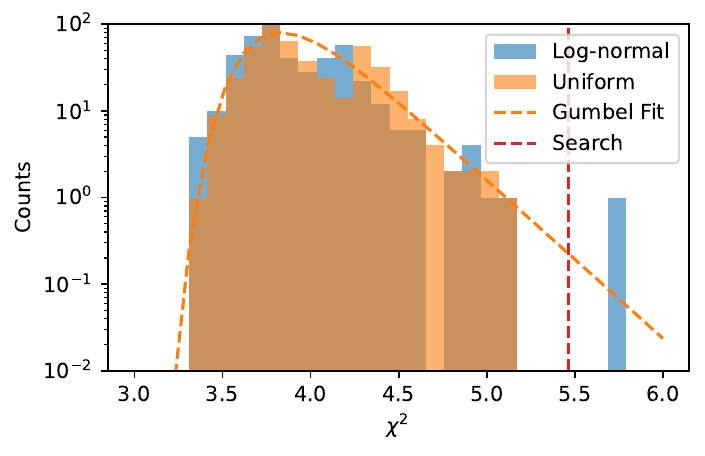}
    \caption{The search (red dashed line), log-normal and uniform monte-carlo simulations (blue and orange histograms correspondingly) results for MJD 59310, with a fit to the uniform monte-carlo simulation (orange solid line).
    We can see that the search result is larger than all uniform monte-carlo samples, but not much larger as demonstrated by the fit to a Gumbel distribution.}
    \label{fig:59310_hist}
\end{figure}

\begin{figure*}[]
    \centering
    \plottwo{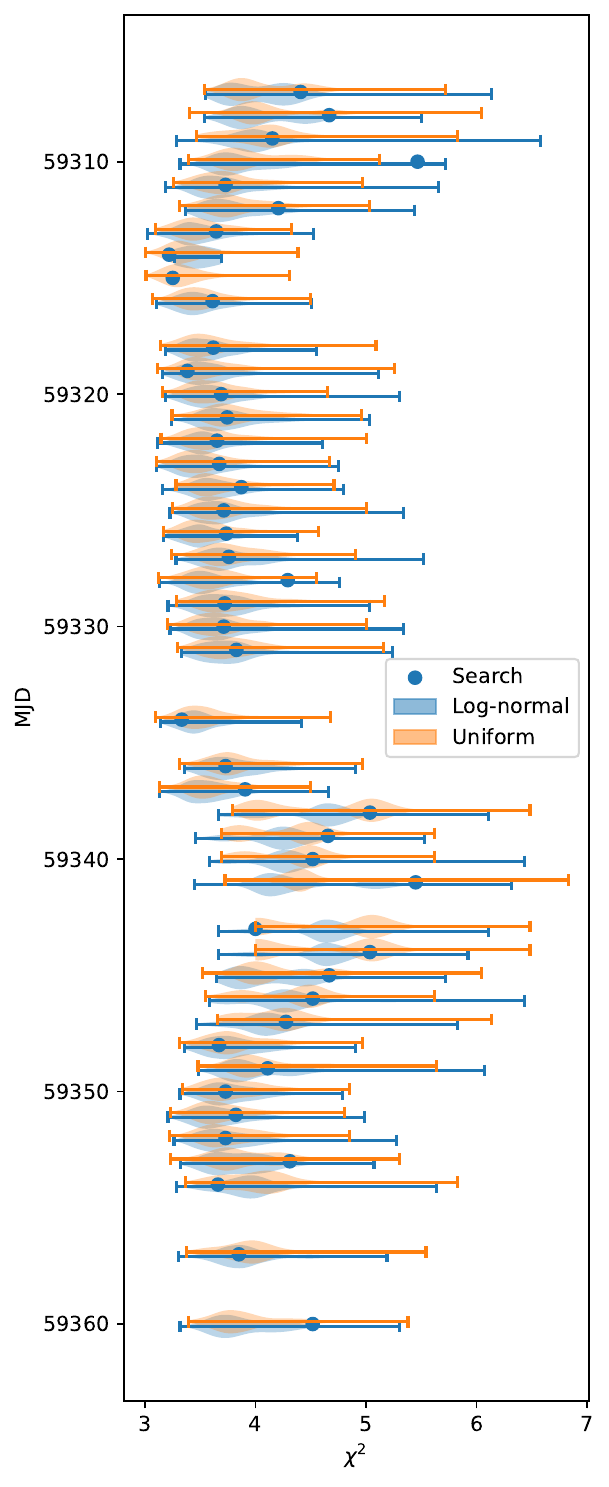}{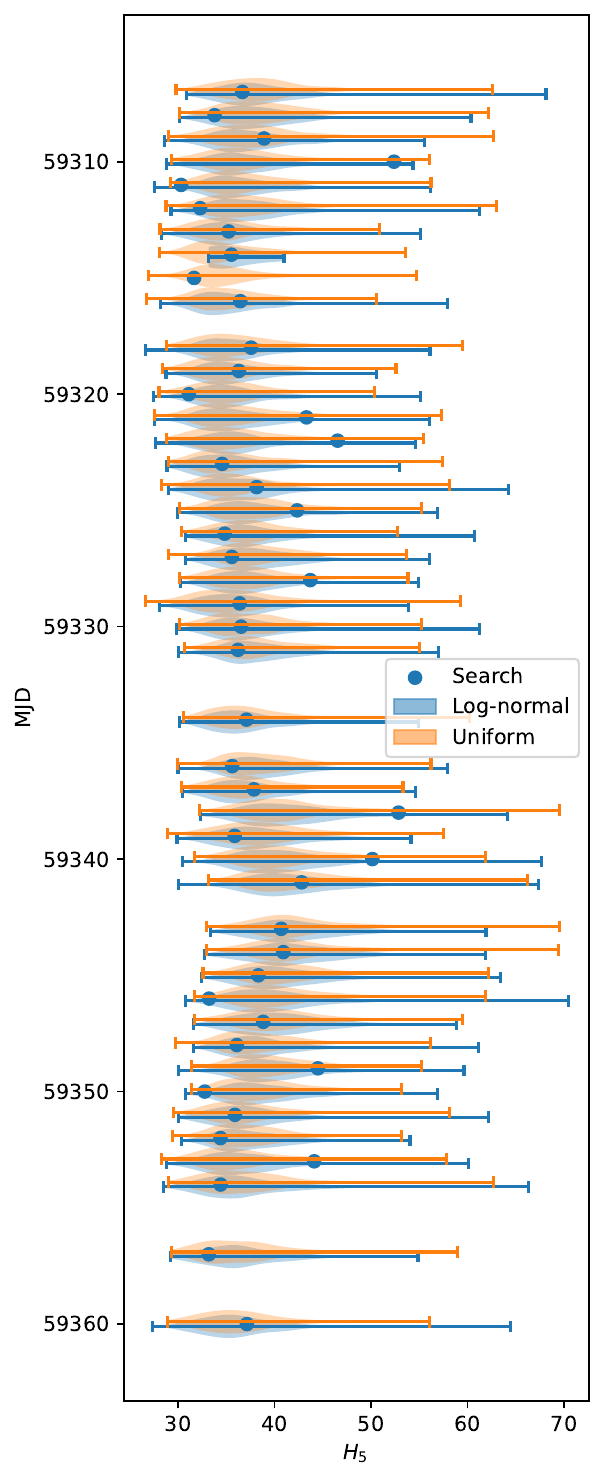}
    \caption{A summary of search (blue dots), log-normal and uniform monte-carlo simulations (blue and orange violins correspondingly) results, using $\chi^2$ (left) and $H_5$ (right) statistics.
    We can see only MJD 59310 marginally passes the log-normal monte-carlo simulation, and all other days are clearly not significant.}
    \label{fig:violins}
\end{figure*}

The test statistics found for other days, and their sampled distributions can be found in Figure \ref{fig:violins}.

\subsection{Result for MJD 59347}
MJD 59347 is the second day in which \cite{du_second-scale_2025} found periodicity, however, as seen in Figure \ref{fig:violins}, in our search we get $\chi^2=4.3$, in contrast to their $\chi^2=5.8$.
This difference is caused by our rejection of 3 bursts due to short waiting times. In the original data, 2 triplets of bursts are detected in this day. Using the $0.05\text{s}$ waiting time threshold rejects only 1 of the bursts, while using our $0.4\text{s}$ threshold reduces each triplet to its representative burst.
Under this reduction in test statistic value, the search result is well within the sampled distribution of null search results
\begin{equation*}
    \text{p-value}_{59347}\ge \frac{1}{500} \,.
\end{equation*}

As we demonstrate in Appendix \ref{app:stat_in_short} for the $H$-statistic, the null distribution of our test greatly changes, especially in its tail, when including bursts with short waiting times.

\section{Discussion}
\subsection{Relation to \cite{du_second-scale_2025}}
We have a few disagreements with the methods implemented by \cite{du_second-scale_2025}, in which a similar periodicity search was done:
\begin{enumerate}
    \item Time of arrival pre-processing - the threshold on waiting times used here is aimed to reject bursts from the short mode of waiting times from the analysis, motivated by an assumption this mode is some internal structure of the bursts, and isn't related to a possible activity period of the engine. We also demonstrate this choice's effect on the test distribution in Appendix \ref{app:stat_in_short}.
    \item Frequency spacing - the spacing in frequency here is defined from the basic principle of limited phase resolution and its implication on the required frequency resolution.
    \item Test statistic - we both retain compatibility by using the $\chi^2$ statistic, but employ also the $H$ statistic, mostly because it doesn't require binning. We also limit the maximal harmonic used in the $H$ test, and argue this is required due to the low number of bursts analysed in each day.
    \item Significance determination - we do not use the commonly used distributions of the $\chi^2$ and $H$ tests, because the approximations used to derive or estimate them are not valid in our case of small number of bursts. We sample mock TOA series and search for periodicity in them, and compare the best test statistics from those searches to the ones found when searching over the FAST data.
\end{enumerate}
Out of those differences, the time of arrival pre-processing is the one tipping the scale and leading us not to discover the periodicity claimed by \cite{du_second-scale_2025}.

\subsection{Rules of thumb}
The most important conclusion from this study is that, when determining the significance of a detection, one needs to properly calculate or sample from the null distribution. In this case, the correlations between bursts' arrival times, in the form of bi-modal waiting times distribution, and the process of searching for periodicity and choosing the period with the best score, must be taken into account when estimating a result's significance.
Other considerations, such as the observationally motivated waiting time threshold, the proper frequency spacing for the search, the smooth $H$-statistic compared to the discrete $\chi^2$-statistic, and the balance between the number of observations and (actual or effective) bins, are also important.

\section{Conclusions}
In this study, We searched for periodicity in the highly active episode of FRB 20201124A, as recorded using FAST, between April and June 2021. The search resulted in the non-detection of strict periodicity in the bursts' arrival times within individual days.

We compared my search strategy and significance determination scheme to the recent works of \cite{du_thorough_2024, du_second-scale_2025}, and claim their detection of periodicity in MJDs 59310, 59347 results from improper significance determination scheme, and discrepancy between waiting time threshold and background TOA sampling.

\section*{Acknowledgements}
DG thanks Eli Waxman, Boaz Katz, Pravir Kumar, Eran Ofek and Sahar Shahaf for useful discussions, and Chen Du for open sharing of his analysis methods and comments.
This research was Supported by the Minerva Foundation with funding from the Federal German Ministry for Education and Research. This research was partially supported by the Israeli Council for Higher Education (CHE) via the Weizmann Data Science Research Center.

\appendix

\section{Statistics in the presence of short waiting time distribution}
\label{app:stat_in_short}
\begin{figure}
    \centering
    \plotone{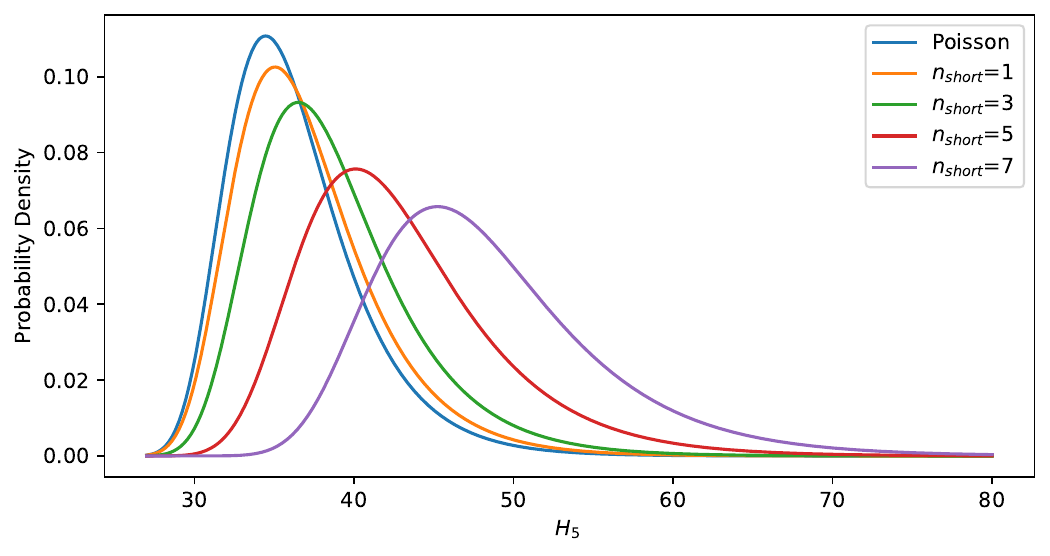}
    \caption{The maximal $H_5$-statistic distribution in a periodicity search for various waiting time distributions. Sets of 30 TOAs are sampled using FRB 20201124A's waiting time distribution constrained to $n_{\text{short}}$ waiting times sampled from the short mode, or assuming a Poisson process. Each is searched for periodicity and the best $H_5$ values are fit as maximum of exponential random variables.
    We can see that larger $H_5$ values are easy to achieve with more waiting times sampled from the short mode.}
    \label{fig:bimodal}
\end{figure}

In \ref{subsubsec:preprocess} we set a threshold on the waiting times used in the periodicity search, this was motivated by the hypothesis that the short waiting times, shorter than the periods we search for, are caused by some sub-structure of the engine (for example the beam shape of pulsars).
Here we also argue it biases the statistical tests we use. For the $\chi^2$-statistic, we can understand this either by having a pair of bursts that will always fall in the same bin, or, for slightly larger separation, as a pair of bursts that contribute to adjacent bins, which when scanning the frequency traverse the phase together and make it easier to align many bursts in a few bins.
For the $H$ test the intuition is similar, but we also performed numerical tests. We sampled $30$ waiting times from the rFRB's waiting time distribution as presented in Figure \ref{fig:wait_times}, with various numbers of waiting times from the short mode, over each we searched for periodicity, and fit the maxima to the distribution of maximum of exponential independent random variables. As a reference we also sample exponential waiting times with the mean burst rate, which corresponds to a Poisson process.
The results are presented in Figure \ref{fig:bimodal}, showing larger values of $H_5$ are achieved the more waiting times are sampled from the short mode of the distribution, as expected.

\bibliography{references}{}

\begin{thebibliography}{}
\expandafter\ifx\csname natexlab\endcsname\relax\def\natexlab#1{#1}\fi
\providecommand{\url}[1]{\href{#1}{#1}}
\providecommand{\dodoi}[1]{doi:~\href{http://doi.org/#1}{\nolinkurl{#1}}}
\providecommand{\doeprint}[1]{\href{http://ascl.net/#1}{\nolinkurl{http://ascl.net/#1}}}
\providecommand{\doarXiv}[1]{\href{https://arxiv.org/abs/#1}{\nolinkurl{https://arxiv.org/abs/#1}}}

\bibitem[{C.~D. Bochenek {et~al.}(2020{\natexlab{a}})Bochenek, McKenna, Belov, Kocz, Kulkarni, Lamb, Ravi, \& Woody}]{bochenek_stare2_2020}
Bochenek, C.~D., McKenna, D.~L., Belov, K.~V., {et~al.} 2020{\natexlab{a}}, \bibinfo{title}{{STARE2}: {Detecting} {Fast} {Radio} {Bursts} in the {Milky} {Way},} Publications of the Astronomical Society of the Pacific, 132, 034202, \dodoi{10.1088/1538-3873/ab63b3}

\bibitem[{C.~D. Bochenek {et~al.}(2020{\natexlab{b}})Bochenek, Ravi, Belov, Hallinan, Kocz, Kulkarni, \& McKenna}]{bochenek_fast_2020}
Bochenek, C.~D., Ravi, V., Belov, K.~V., {et~al.} 2020{\natexlab{b}}, \bibinfo{title}{A fast radio burst associated with a {Galactic} magnetar,} Nature, 587, 59, \dodoi{10.1038/s41586-020-2872-x}

\bibitem[{ {Chime/Frb Collaboration} {et~al.}(2020){Chime/Frb Collaboration}, Amiri, Andersen, Bandura, Bhardwaj, Boyle, Brar, Chawla, Chen, Cliche, Cubranic, Deng, Denman, Dobbs, Dong, Fandino, Fonseca, Gaensler, Giri, Good, Halpern, Hessels, Hill, Höfer, Josephy, Kania, Karuppusamy, Kaspi, Keimpema, Kirsten, Landecker, Lang, Leung, Li, Lin, Marcote, Masui, McKinven, Mena-Parra, Merryfield, Michilli, Milutinovic, Mirhosseini, Naidu, Newburgh, Ng, Nimmo, Paragi, Patel, Pen, Pinsonneault-Marotte, Pleunis, Rafiei-Ravandi, Rahman, Ransom, Renard, Sanghavi, Scholz, Shaw, Shin, Siegel, Singh, Smegal, Smith, Stairs, Tendulkar, Tretyakov, Vanderlinde, Wang, Wang, Wulf, Yadav, \& Zwaniga}]{chimefrb_collaboration_periodic_2020}
{Chime/Frb Collaboration}, Amiri, M., Andersen, B.~C., {et~al.} 2020, \bibinfo{title}{Periodic activity from a fast radio burst source,} Nature, 582, 351, \dodoi{10.1038/s41586-020-2398-2}

\bibitem[{ {CHIME/FRB Collaboration} {et~al.}(2020){CHIME/FRB Collaboration}, Andersen, Bandura, Bhardwaj, Bij, Boyce, Boyle, Brar, Cassanelli, Chawla, Chen, Cliche, Cook, Cubranic, Curtin, Denman, Dobbs, Dong, Fandino, Fonseca, Gaensler, Giri, Good, Halpern, Hill, Hinshaw, Höfer, Josephy, Kania, Kaspi, Landecker, Leung, Li, Lin, Masui, McKinven, Mena-Parra, Merryfield, Meyers, Michilli, Milutinovic, Mirhosseini, Münchmeyer, Naidu, Newburgh, Ng, Patel, Pen, Pinsonneault-Marotte, Pleunis, Quine, Rafiei-Ravandi, Rahman, Ransom, Renard, Sanghavi, Scholz, Shaw, Shin, Siegel, Singh, Smegal, Smith, Stairs, Tan, Tendulkar, Tretyakov, Vanderlinde, Wang, Wulf, \& Zwaniga}]{chimefrb_collaboration_bright_2020}
{CHIME/FRB Collaboration}, Andersen, B.~C., Bandura, K.~M., {et~al.} 2020, \bibinfo{title}{A bright millisecond-duration radio burst from a {Galactic} magnetar,} Nature, 587, 54, \dodoi{10.1038/s41586-020-2863-y}

\bibitem[{M. Cruces {et~al.}(2021)Cruces, Spitler, Scholz, Lynch, Seymour, Hessels, Gouiffés, Hilmarsson, Kramer, \& Munjal}]{cruces_repeating_2021}
Cruces, M., Spitler, L.~G., Scholz, P., {et~al.} 2021, \bibinfo{title}{Repeating behaviour of {FRB} 121102: periodicity, waiting times, and energy distribution,} Monthly Notices of the Royal Astronomical Society, 500, 448, \dodoi{10.1093/mnras/staa3223}

\bibitem[{O.~C. de~Jager {et~al.}(1989)de~Jager, Raubenheimer, \& Swanepoel}]{de_jager_powerful_1989}
de~Jager, O.~C., Raubenheimer, B.~C., \& Swanepoel, J. W.~H. 1989, \bibinfo{title}{A powerful test for weak periodic signals with unknown light curve shape in sparse data.,} Astronomy and Astrophysics, 221, 180.
\newblock \url{https://ui.adsabs.harvard.edu/abs/1989A&A...221..180D}

\bibitem[{C. Du {et~al.}(2024)Du, Huang, Zhang, Rodin, Fedorova, Kurban, \& Li}]{du_thorough_2024}
Du, C., Huang, Y.-F., Zhang, Z.-B., {et~al.} 2024, \bibinfo{title}{A {Thorough} {Search} for {Short}-timescale {Periodicity} in {Four} {Active} {Repeating} {Fast} {Radio} {Bursts},} The Astrophysical Journal, 977, 129, \dodoi{10.3847/1538-4357/ad8cd5}

\bibitem[{C. Du {et~al.}(2025)Du, Huang, Geng, Gao, Zhang, Deng, Cui, Liao, Jiang, Zhang, Wang, Hu, Dong, Xu, Li, Zou, \& Kurban}]{du_second-scale_2025}
Du, C., Huang, Y.-F., Geng, J.-J., {et~al.} 2025, \bibinfo{title}{A second-scale periodicity in an active repeating fast radio burst source,} arXiv, \dodoi{10.48550/arXiv.2503.12013}

\bibitem[{Y. Feng {et~al.}(2023)Feng, Li, Zhang, Tsai, Wang, Yang, Qu, Wang, Zhou, Niu, Miao, Yuan, Xu, Lynch, Armentrout, Gregory, Meng, Wang, Chen, Dai, Niu, Xue, Yao, Zhang, Zhang, Zhu, \& Zhu}]{feng_extreme_2023}
Feng, Y., Li, D., Zhang, Y.-K., {et~al.} 2023, \bibinfo{title}{An extreme active repeating fast radio burst in a clean environment,} arXiv, \dodoi{10.48550/arXiv.2304.14671}

\bibitem[{D. Gazith {et~al.}(2025)Gazith, Pearlman, \& Zackay}]{gazith_recovering_2025}
Gazith, D., Pearlman, A.~B., \& Zackay, B. 2025, \bibinfo{title}{Recovering {Pulsar} {Periodicity} from {Time}-of-arrival {Data} by {Finding} the {Shortest} {Vector} in a {Lattice},} The Astrophysical Journal, 979, 48, \dodoi{10.3847/1538-4357/ad9449}

\bibitem[{F. Kirsten {et~al.}(2021)Kirsten, Snelders, Jenkins, Nimmo, van~den Eijnden, Hessels, Gawroński, \& Yang}]{kirsten_detection_2021}
Kirsten, F., Snelders, M.~P., Jenkins, M., {et~al.} 2021, \bibinfo{title}{Detection of two bright radio bursts from magnetar {SGR} 1935 + 2154,} Nature Astronomy, 5, 414, \dodoi{10.1038/s41550-020-01246-3}

\bibitem[{F. Kirsten {et~al.}(2022)Kirsten, Marcote, Nimmo, Hessels, Bhardwaj, Tendulkar, Keimpema, Yang, Snelders, Scholz, Pearlman, Law, Peters, Giroletti, Paragi, Bassa, Hewitt, Bach, Bezrukovs, Burgay, Buttaccio, Conway, Corongiu, Feiler, Forssén, Gawroński, Karuppusamy, Kharinov, Lindqvist, Maccaferri, Melnikov, Ould-Boukattine, Possenti, Surcis, Wang, Yuan, Aggarwal, Anna-Thomas, Bower, Blaauw, Burke-Spolaor, Cassanelli, Clarke, Fonseca, Gaensler, Gopinath, Kaspi, Kassim, Lazio, Leung, Li, Lin, Masui, Mckinven, Michilli, Mikhailov, Ng, Orbidans, Pen, Petroff, Rahman, Ransom, Shin, Smith, Stairs, \& Vlemmings}]{kirsten_repeating_2022}
Kirsten, F., Marcote, B., Nimmo, K., {et~al.} 2022, \bibinfo{title}{A repeating fast radio burst source in a globular cluster,} Nature, 602, 585, \dodoi{10.1038/s41586-021-04354-w}

\bibitem[{D.~R. Lorimer {et~al.}(2007)Lorimer, Bailes, McLaughlin, Narkevic, \& Crawford}]{lorimer_bright_2007}
Lorimer, D.~R., Bailes, M., McLaughlin, M.~A., Narkevic, D.~J., \& Crawford, F. 2007, \bibinfo{title}{A bright millisecond radio burst of extragalactic origin,} Science, 318, 777

\bibitem[{B. Marcote {et~al.}(2017)Marcote, Paragi, Hessels, Keimpema, van Langevelde, Huang, Bassa, Bogdanov, Bower, Burke-Spolaor, Butler, Campbell, Chatterjee, Cordes, Demorest, Garrett, Ghosh, Kaspi, Law, Lazio, McLaughlin, Ransom, Salter, Scholz, Seymour, Siemion, Spitler, Tendulkar, \& Wharton}]{marcote_repeating_2017}
Marcote, B., Paragi, Z., Hessels, J. W.~T., {et~al.} 2017, \bibinfo{title}{The {Repeating} {Fast} {Radio} {Burst} {FRB} 121102 as {Seen} on {Milliarcsecond} {Angular} {Scales},} The Astrophysical Journal, 834, L8, \dodoi{10.3847/2041-8213/834/2/L8}

\bibitem[{B. Marcote {et~al.}(2020)Marcote, Nimmo, Hessels, Tendulkar, Bassa, Paragi, Keimpema, Bhardwaj, Karuppusamy, Kaspi, Law, Michilli, Aggarwal, Andersen, Archibald, Bandura, Bower, Boyle, Brar, Burke-Spolaor, Butler, Cassanelli, Chawla, Demorest, Dobbs, Fonseca, Giri, Good, Gourdji, Josephy, Kirichenko, Kirsten, Landecker, Lang, Lazio, Li, Lin, Linford, Masui, Mena-Parra, Naidu, Ng, Patel, Pen, Pleunis, Rafiei-Ravandi, Rahman, Renard, Scholz, Siegel, Smith, Stairs, Vanderlinde, \& Zwaniga}]{marcote_repeating_2020}
Marcote, B., Nimmo, K., Hessels, J. W.~T., {et~al.} 2020, \bibinfo{title}{A repeating fast radio burst source localized to a nearby spiral galaxy,} Nature, 577, 190, \dodoi{10.1038/s41586-019-1866-z}

\bibitem[{C.~H. Niu {et~al.}(2022)Niu, Aggarwal, Li, Zhang, Chatterjee, Tsai, Yu, Law, Burke-Spolaor, Cordes, Zhang, Ocker, Yao, Wang, Feng, Niino, Bochenek, Cruces, Connor, Jiang, Dai, Luo, Li, Miao, Niu, Anna-Thomas, Sydnor, Stern, Wang, Yuan, Yue, Zhou, Yan, Zhu, \& Zhang}]{niu_repeating_2022}
Niu, C.~H., Aggarwal, K., Li, D., {et~al.} 2022, \bibinfo{title}{A repeating fast radio burst associated with a persistent radio source,} Nature, 606, 873, \dodoi{10.1038/s41586-022-04755-5}

\bibitem[{J.-R. Niu {et~al.}(2022)Niu, Zhu, Zhang, Yuan, Zhou, Zhang, Jiang, Han, Li, Lee, Wang, Feng, Li, Luo, Wang, Dai, Miao, Niu, Xu, Zhang, Wang, Wang, \& Xu}]{niu_fast_2022}
Niu, J.-R., Zhu, W.-W., Zhang, B., {et~al.} 2022, \bibinfo{title}{{FAST} {Observations} of an {Extremely} {Active} {Episode} of {FRB} {20201124A}. {IV}. {Spin} {Period} {Search},} Research in Astronomy and Astrophysics, 22, 124004, \dodoi{10.1088/1674-4527/ac995d}

\bibitem[{K.~M. Rajwade {et~al.}(2020)Rajwade, Mickaliger, Stappers, Morello, Agarwal, Bassa, Breton, Caleb, Karastergiou, Keane, \& Lorimer}]{rajwade_possible_2020}
Rajwade, K.~M., Mickaliger, M.~B., Stappers, B.~W., {et~al.} 2020, \bibinfo{title}{Possible periodic activity in the repeating {FRB} 121102,} Monthly Notices of the Royal Astronomical Society, 495, 3551, \dodoi{10.1093/mnras/staa1237}

\bibitem[{P. Scholz \&  {Chime/Frb Collaboration}(2020)Scholz \& {Chime/Frb Collaboration}}]{scholz_bright_2020}
Scholz, P., \& {Chime/Frb Collaboration}. 2020, \bibinfo{title}{A bright millisecond-timescale radio burst from the direction of the {Galactic} magnetar {SGR} 1935+2154,} The Astronomer's Telegram, 13681, 1.
\newblock \url{https://ui.adsabs.harvard.edu/abs/2020ATel13681....1S}

\bibitem[{S.~P. Tendulkar {et~al.}(2017)Tendulkar, Bassa, Cordes, Bower, Law, Chatterjee, Adams, Bogdanov, Burke-Spolaor, Butler, Demorest, Hessels, Kaspi, Lazio, Maddox, Marcote, McLaughlin, Paragi, Ransom, Scholz, Seymour, Spitler, van Langevelde, \& Wharton}]{tendulkar_host_2017}
Tendulkar, S.~P., Bassa, C.~G., Cordes, J.~M., {et~al.} 2017, \bibinfo{title}{The {Host} {Galaxy} and {Redshift} of the {Repeating} {Fast} {Radio} {Burst} {FRB} 121102,} The Astrophysical Journal, 834, L7, \dodoi{10.3847/2041-8213/834/2/L7}

\bibitem[{E. Waxman(2017)Waxman}]{waxman_origin_2017}
Waxman, E. 2017, \bibinfo{title}{On the {Origin} of {Fast} {Radio} {Bursts} ({FRBs}),} The Astrophysical Journal, 842, 34, \dodoi{10.3847/1538-4357/aa713e}

\bibitem[{H. Xu {et~al.}(2022)Xu, Niu, Chen, Lee, Zhu, Dong, Zhang, Jiang, Wang, Xu, Zhang, Fu, Filippenko, Peng, Zhou, Zhang, Wang, Feng, Li, Brink, Li, Lu, Yang, Caballero, Cai, Chen, Dai, Djorgovski, Esamdin, Gan, Guhathakurta, Han, Hao, Huang, Jiang, Li, Li, Li, Li, Li, Liu, Luo, Men, Niu, Peng, Qian, Song, Stern, Stockton, Sun, Wang, Wang, Wang, Wang, Wu, Xiao, Xiong, Xu, Xu, Yang, Yang, Yao, Yi, Yue, Yu, Yu, Yuan, Zhang, Zhang, Zhang, Zhao, Zheng, Zhu, \& Zou}]{xu_fast_2022}
Xu, H., Niu, J.~R., Chen, P., {et~al.} 2022, \bibinfo{title}{A fast radio burst source at a complex magnetized site in a barred galaxy,} Nature, 609, 685, \dodoi{10.1038/s41586-022-05071-8}

\bibitem[{C.~F. Zhang {et~al.}(2020)Zhang, Jiang, Men, Wang, Xu, Xu, Niu, Zhou, Guan, Han, Jiang, Lee, Li, Lin, Niu, Wang, Wang, Xu, Yu, Zhang, \& Zhu}]{zhang_highly_2020}
Zhang, C.~F., Jiang, J.~C., Men, Y.~P., {et~al.} 2020, \bibinfo{title}{A highly polarised radio burst detected from {SGR} 1935+2154 by {FAST},} The Astronomer's Telegram, 13699, 1.
\newblock \url{https://ui.adsabs.harvard.edu/abs/2020ATel13699....1Z}

\end{thebibliography}
\bibliographystyle{aasjournalv7}

\end{document}